\documentstyle[12pt,amstex,righttag]{article}
\begin{document}

\vspace*{-4cm}

\ \\[1mm]

\begin{center}
{\large\bf Generalization of the Lie-Trotter Product Formula for
$q$-Exponential Operators}

\ \\[1mm]

{\sl A.K. Rajagopal}

Naval Research Laboratory\\[-1mm]
Washington DC 20375-5320, USA\\[-1 mm]

and

 {\sl Constantino Tsallis} 

Centro Brasileiro de Pesquisas F\'{\i}sicas - CBPF/CNPq\\[-1mm]
Rua Dr. Xavier Sigaud, 150 \\[-1mm]
22290-180 - Rio de Janeiro, RJ, Brazil \\[-1 mm]
\end{center}

\begin{abstract}
The Lie-Trotter formula $e^{\hat{A}+\hat{B}} = \lim_{N\rightarrow \infty} \left(e^{\hat{A}/N}
e^{\hat{B}/N}\right)^N \ $ is of great utility in a variety of quantum problems ranging from the theory of path integrals and Monte Carlo methods in theoretical chemistry, to many-body and thermostatistical calculations. We generalize it for the $q$-exponential function $ e_q (x) = [1+ (1-q) x]^{(1/(1-q))} \ $ (with $e_1(x)=e^x$), and prove $ e_q(\hat{A}+\hat{B}+(1-q)
[\hat{A}\hat{B}+\hat{B}\hat{A}] /2) =
\lim_{N\rightarrow \infty}
\left\{\left[e_{1-(1-q)N}\left(\hat{A}/N\right)\right] 
\left[e_{1-(1-q)N}\left(\hat{B}/N\right)\right]\right\}^N$. This extended formula is expected to be similarly useful in the nonextensive situations.\\

\end{abstract}

The Lie-Trotter product formula states that if $\hat{A}$ and  
$\hat{B}$ are  not necessarily commuting finite square matrices or bounded
operators with respect to some convenient norm, then
\begin{equation}
e^{\hat{A}+\hat{B}} = \lim_{N\rightarrow \infty} \left(e^{\hat{A}/N}
e^{\hat{B}/N}\right)^N \ .
\end{equation} 
This formula has been central in the development of path integral
approaches to quantum theory, stochastic theory, and in quantum
statistical mechanics \cite{1,2}. In particular, Suzuki has employed
this to develop quantum statistical Monte Carlo methods \cite{3,straub},
general theory of path integrals with application to many-body
theories and statistical physics \cite{4}, and more recently to
mathematical physics \cite{5}. In the past decade, the use of
monomial form of the exponential function \cite{6} defined by
(hereafter called $q$-exponential)
\begin{equation}
e_q (x) = [1+ (1-q) x]^{1/(1-q)} \;\;\;\;\;(q \in \cal{R})\,
\end{equation} 

\vspace{0.1cm}
\noindent
with a cut-off, for $q<1$, when $[1+ (1-q) x] < 0$ (analogously, for $q>1$, $e_q(x)$ diverges at $x=1/(q-1)$), has played an important role in the development of nonextensive
statistical physics \cite{7}. Here we consider
the corresponding $q$-exponential operator defined similarly by
\begin{equation}
e_q (\hat{A}) = [1+(1-q)\hat{A}]^{1/(1-q)}\;\;\;\;\;\; (q \in \cal{R})      \ .
\end{equation}
Here, $\hat{A}$ can be a $c$-number, a square matrix,
or an operator. For $q$ going to $1$, this goes to the usual
exponential form defined in Eq. (1). There is a large class of
problems (Levy-like\cite{levy} and correlated-like\cite{bukman} anomalous diffusion, turbulence\cite{boghosian} in nonneutral plasma, nonlinear dynamics\cite{marcelo}, solar neutrino problem\cite{quarati}, cosmology\cite{cosmos}, electron-positron collisions\cite{bediaga}, reassociation of heme-ligands in folded proteins\cite{bemski}, among many others) with a (multi)fractal structure in the relevant space-time in which this and related functional forms emerge naturally. Many of these have received until now only a classical approach; however, their quantum counterparts could be analyzed as well, and then the operator form we are addressing would be the natural one to employ. In this paper, we
wish to present the generalization of the formula in Eq. (1) for
these monomial functions in the form:
\begin{equation}
e_q(\hat{A}+\hat{B}+(1-q)\{\hat{A},\hat{B}\}/2) =
\lim_{N\rightarrow \infty}\left\{\left[e_{1-(1-q)N}\left(\hat{A}/N\right)\right] 
\left[e_{1-(1-q)N}\left(\hat{B}/N\right)\right]\right\}^N 
\end{equation}
where $\{\hat{A},\hat{B}\} \equiv (\hat{A}\hat{B}+\hat{B}\hat{A})$. 
Note that $e_{1-(1-q)N}(\hat{A}/N) = [1+(1-q)\hat{A}]^{(1/(1-q)N)}$,
has the  properties that it goes to $e^{(\hat{A}/N)}$ both when $q=1$
for finite $N$ and for $q\neq 1$, but $N\rightarrow \infty$, so that
we recover Eq. (1) appropriately. To establish Eq. (4), we first
consider the following pair of operators:
\begin{eqnarray}
&& \hat{C} = \left[e_{\tilde{q}}\left(\frac{\hat{A}+\hat{B}+(1-\tilde{q})\{\hat{A},\hat{B}\}/2N}
{N}\right)\right]; \nonumber \\
&& \hat{D} = e_{\tilde{q}}(\hat{A}/N)e_{\tilde{q}}(\hat{B}/N) \ .   
\end{eqnarray}

Then, following similar steps as in the case of the exponential
operators in \cite{1}, the norms of these operators are found to be
bounded by the   following inequalities:
\begin{eqnarray}
&\|\hat{C}\| \leq \left[e_{\tilde{q}}\left(\left\|\frac{\hat{A}+\hat{B}
+ (1-\tilde{q})\{\hat{A},\hat{B}\}/2N}{N}\right\|\right)\right] \leq
\left[e_{\tilde{q}}\left(\frac{\|\hat{A}\| + \|\hat{B}\| +
\frac{(1-\tilde{q})}{N} \|\hat{A}\|\|\hat{B}\|}{N}\right)\right]
\end{eqnarray}
and
\begin{eqnarray}
&& \|\hat{D}\| \leq \|e_{\tilde{q}}(\hat{A}/N)\|\|e_{\tilde{q}}(\hat{B}/N)\|
\nonumber \\
&& \hspace{0.9cm} \leq e_{\tilde{q}}(\|\hat{A}\|/N)e_{\tilde{q}} (\|\hat{B}\|/N)
\nonumber \\
&& \hspace{0.9cm} = e_{\tilde{q}}\left(\left(\|\hat{A}\| + \|\hat{B}\| +
\frac{(1-\tilde{q})}{N} \|\hat{A}\|\|\hat{B}\|\right)/N\right)
\nonumber  \\
&& \hspace{0.9cm} = \left[1+ \frac{(1-\tilde{q})}{N}\left(\|\hat{A}\| + \|\hat{B}\| +
\frac{(1-\tilde{q})}{N} \|\hat{A}\|\|\hat{B}\|\right)\right]^{(1/(1-\tilde{q}))}
\nonumber \\
&&  \hspace{0.9cm} =\left[e_{\tilde{q}}\left(\frac{\|\hat{A}\| + \|\hat{B}\| +
\frac{(1-\tilde{q})}{N} \|\hat{A}\|\|\hat{B}\|}{N}\right)\right] 
\end{eqnarray}
where we have used that (i) the norm of an increasing function of an operator cannot exceed the function of the norm of that operator, (ii) the norm of a sum of two operators cannot exceed the sum of the norms of those operators, and (iii) the norm of the product of two operators cannot exceed the product of the norms of those operators. These expressions are valid for bounded operators as long as the restrictions mentioned after Eq. (2) are obeyed. In the last line in Eq. (7), we used the identity which follows from
Eq. (2) for $c$-numbers, that
\begin{equation}
e_{\tilde{q}} (x) e_{\tilde{q}} (y) = e_{\tilde{q}} (x+y+(1-\tilde{q})xy)
\end{equation}
in which we set $x=\|\hat{A}\|/N$, and $y=\|\hat{B}\|/N$. Thus
the norms of the two operators are found to be bounded by the same
quantity. We now calculate the norm $\|\hat{C}^N-\hat{D}^N\|$. We
therefore consider the operator identity given in \cite{1,2},
\begin{equation}
\hat{C}^N-\hat{D}^N = \sum^N_{k=1}
\hat{C}^{k-1}(\hat{C}-\hat{D})\hat{D}^{N-k} \ ,
\end{equation}
and find a bound on the left side of this equation by  the norm on
the right hand side, using Eqs. (6) and (7):
\begin{eqnarray}
&& \|\hat{C}^N-\hat{D}^N\| \leq 
\sum^N_{k=1}
\|\hat{C}\|^{k-1}\|(\hat{C}-\hat{D})\|\|\hat{D}\|^{N-k}\hspace{8.5cm} \
\ 
\nonumber \\
&& \hspace{2.3cm} \leq N\|(\hat{C}-\hat{D})\|
\left[1+\frac{(1-\tilde{q})}{N} \left(\|\hat{A}\|+\|\hat{B}\| 
+ \frac{(1-\tilde{q})}{N}
\|\hat{A}\|\|\hat{B}\|\right)\right]^{((N-1)/(1-\tilde{q}))}
\end{eqnarray}
Expanding $\hat{C}$ and $\hat{D}$ in a power series in $(1/N)$, we
now estimate the norm of their difference for large $N$

$
\parallel (\hat{C}-\hat{D})\parallel $ $\leq \left|\left| 
\begin{array}{l}
\left( 
\begin{array}{l}
1+\left[ \frac{\hat{A}+\hat{B}+(1-\tilde{q})\left\{ \hat{A},\hat{B}\right\}
/2N}N\right]  \\ 
+\frac{\bar{q}}2\left[ \frac{\hat{A}+\hat{B}+(1-\tilde{q})\left\{ \hat{A},%
\hat{B}\right\} /2N}N\right] ^2+...
\end{array}
\right)  \\ 
-\left( 1+\left[ \frac{\hat{A}}N\right] +\frac{\bar{q}}2\left[ \frac{\hat{A}}%
N\right] \Sp 2 \\  \endSp +...\right) \left( 1+\left[ \frac{\hat{B}}N\right]
+\frac{\bar{q}}2\left[ \frac{\hat{B}}N\right] ^2+...\right) 
\end{array}
\right|\right| $

\begin{equation}
\leq \frac 1{2N^2}\parallel\! \left[ \hat{A},\hat{B}\right]\!     +o(\frac{1-\tilde{q}}{N^3})      \parallel ,%
\hspace{6.6cm}\ 
\end{equation}

\noindent
where $[\hat{A},\hat{B}] \equiv \hat{A}\hat{B}-\hat{B}\hat{A}$. It is worth noticing that there is no $1/N$ term.\\
By using inequality (11) into (10) we have thus obtained the following inequality for any $\tilde{q}$ and large  $N$:
\begin{equation}
\|\hat{C}^N - \hat{D}^N\|\leq \frac{1}{2N} \|[\hat{A},\hat{B}]\|
\left[1+\frac{(1-\tilde{q})}{N} \left(\|\hat{A}\|+
\|\hat{B}\| + \frac{(1-\tilde{q})}{N} \|\hat{A}\|
\|\hat{B}\|\right)\right]^{((N-1)/(1-\tilde{q}))}\ .
\end{equation}
We now let $\tilde{q}=1-(1-q)N$, which implies  $ o(\frac{1-\tilde{q}}{N^3})=o(1/N^2)$,  and take the limit $N\rightarrow
\infty$, thus showing that the right side of Eq. (12) tends to zero, which implies
the required generalization of the Lie-Trotter formula, Eq. (4), i.e., 
\begin{equation}
\{1+(1-q)[\hat{A}+\hat{B}+(1-q)\{\hat{A},\hat{B}\}/2]\}^{\frac{1}{1-q}}
=lim_{N \rightarrow \infty}
\{[1+(1-q) \hat{A}]^{\frac{1}{(1-q)N}} [1+(1-q) \hat{B}]^{\frac{1}{(1-q)N}}\}^N
\end{equation}

In conclusion, we have generalized here (Eq. (4) or, equivalently, Eq. (13)) the Lie-Trotter formula, for $q$-exponential operators which occur in the development of nonextensive statistical physics\cite{7}. It is expected that this formula will be useful in quantum versions of generalized simulated annealing\cite{GSA} and in accelerating the rate of convergence of algorithms used in quantum chemistry\cite{straub}, as well as for performing calculations of quantum nonextensive systems.
\vspace{0.5cm}

We are grateful to A.R. Plastino and S. Abe for useful remarks, and to J.E. Straub for communicating to us reference \cite{straub} prior to its publication. AKR acknowledges the U.S. Office of Naval Research for partial support of his work, and CT acknowledges partial support by CNPq and PRONEX/FINEP (Brazilian agencies).


\begin{thebibliography}{99}
\bibitem{1} G. Roepstorff, {\it Path Integral Approach to Quantum
Physics}, Pp.-54-56, Springer-Verlag, New York (1994).
\bibitem{2} M. Reed and B. Simon, {\it Methods of Modern Mathematical
Physics}, Vol. I, Pp. 205-296, Academic Press, New York (1972).
\bibitem{3} M. Suzuki, J. Stat. Phys. {\bf 43}, 883 (1986).
\bibitem{straub}J.E. Straub and T. Whitfield, preprint (1998).
\bibitem{4} M. Suzuki, J. Math. {\bf 32}, 400 (1991).
\bibitem{5} M. Suzuki, Comm. Math. Phys. {\bf 163}, 491 (1994).
\bibitem{6} C. Tsallis, Qu\'{\i}mica Nova (Brazil) {\bf 17}, 468
(1994); E.P. Borges, J. Phys. {\bf A31}, 5281 (1998). 
\bibitem{7} C. Tsallis, J. Stat. Phys. {\bf 52}, 479 (1988); C. Tsallis, R.S. Mendes and A.R. Plastino, Physica A {\bf 261}, 534 (1998). A regularly updated bibliography is accessible at http://tsallis.cat.cbpf.br/biblio.htm
\bibitem{levy}C. Tsallis, S.V.F. Levy, A.M.C. de Souza and R. Maynard, Phys. Rev. Lett. {\bf 75}, 3589 (1995) [Erratum: {\bf 77}, 5442 (1996)].
\bibitem{bukman}C. Tsallis and D.J. Bukman, Phys Rev E {\bf 54}, R2197 (1996).
\bibitem{boghosian}B.M. Boghosian, Phys. Rev. E {\bf 53}, 4754 (1996).
\bibitem{marcelo}M.L. Lyra and C. Tsallis, Phys. Rev. Lett. {\bf 80}, 53 (1998).
\bibitem{quarati}P. Quarati, A. Carbone, G. Cervino, G. Kaniadakis, A. Lavagno and E. Miraldi, Nucl. Phys. A {\bf 621}, 345c (1997).
\bibitem{cosmos}V.H. Hamity and D.E. Barraco, Phys. Rev. Lett. {\bf 76}, 4664 (1996); D.F. Torres, H. Vucetich and A. Plastino, Phys. Rev. Lett. {\bf 79}, 1588 (1997) [Erratum: {\bf 80}, 3889 (1998)].
\bibitem{bediaga}I. Bediaga, E.M.F. Curado and J. Miranda, preprint (1999).
\bibitem{bemski}C. Tsallis, G. Bemski and R.S. Mendes, preprint (1998).
\bibitem{GSA}C. Tsallis and D.A. Stariolo, Physica A {\bf 233}, 395 (1996); I. Andricioaei and J.E. Straub, Phys. Rev. E {\bf 53}, R3055 (1996); P. Serra, A.F. Stanton, S. Kais and R.E. Bleil, J. Chem. Phys. {\bf 106}, 7170 (1997); U.H.E. Hansmann and Y. Okamoto, Phys. Rev. E {\bf 56}, 2228 (1997); M.R. Lemes, C.R Zacharias and A. Dal Pino Jr., Phys. Rev. B {\bf 56}, 9279 (1997). 
\end{thebibliography}
\end{document}